\documentstyle[twocolumn,floats,epsf,aps,prl]{revtex} 

\begin{document}
\draft 
\preprint{Submitted to LT'22} 

\title{Density matrix purification due to continuous quantum
measurement} 

\author{Alexander N. Korotkov} 

\address{Department of Physics, State University of New York,
Stony Brook, NY 11794-3800 } 

%\date{\today}

\maketitle 

\begin{abstract}
We consider the continuous quantum measurement of a two-level system, 
for example, a single-Cooper-pair box measured by a single-electron
transistor or a double-quantum dot measured by a quantum point contact. 
While the approach most commonly used describes the gradual decoherence
of the system due to the measurement, we show that when taking into 
account the detector output, we get the opposite effect: gradual 
purification of the density matrix. The competition between 
purification due to measurement and decoherence due to interaction with 
the environment can be described by a simple Langevin equation which 
couples the random evolution of the system density matrix and the 
stochastic detector output. The gradual density matrix purification 
due to continuous measurement may be verified experimentally 
using present-day technology. The effect can be useful for quantum 
computing. 
\end{abstract}


\narrowtext 

        The active research on quantum computing as well as the 
progress in experimental techniques have motivated renewed 
interest in the problems of quantum measurement, including
the long-standing ``philosophical'' questions. In contrast
to the usual case of averaging over a large ensemble of similar 
quantum systems, it is becoming possible to study experimentally 
the evolution of an individual quantum system. In this paper we consider 
the continuous measurement of a two-level system by a ``weakly responding'' 
\cite{Kor-meas} detector which can be treated as a classical device. 

While after averaging over the ensemble the continuous measurement 
leads to the gradual decoherence of the system density matrix,   
the situation is completely different in the case of an individual
quantum system. In particular, the system evolution becomes dependent
(``conditioned'') on the particular detector output. 
The theory of conditioned evolution of a pure wavefunction was  
developed relatively long ago, mainly for the purposes of quantum optics 
(see, e.g.\, Ref.\ \cite{Plenio} and references therein). 
However, for solid state structures the problem of continuous quantum 
measurement with an account of the measurement result has only been 
addressed recently \cite{Kor-meas}, with the main emphasis on the mixed 
quantum states and the detector nonideality.

        The evolution of the density matrix $\sigma$ of a double-dot 
with the tunneling matrix element $H$ and energy asymmetry $\varepsilon$ 
can be described by nonlinear equations 
        \begin{eqnarray}
&&\dot{\sigma}_{11}=  -\dot{\sigma}_{22}=  -2(H/\hbar )\,\mbox{Im} \sigma_{12}
\nonumber \\
&& \,\,\,\,\,\,\,\,  -\sigma_{11}\sigma_{22}(2\Delta I/S_I) [I(t)-I_0], 
        \label{1}\\
&& {\dot\sigma}_{12}=  i(\varepsilon/\hbar )\sigma_{12}+ i(H/\hbar ) 
(\sigma_{11}-\sigma_{22})
\nonumber \\ 
&&  +( \sigma_{11}-  \sigma_{22}) (\Delta I/S_I) 
[I(t)-I_0] \sigma_{12} -\gamma \sigma_{12},  
        \label{2} \end{eqnarray}
where $I(t)$ is the particular detector output (we assume electric 
current), $I_0=(I_1+I_2)/2$, $I_1$ and $I_2$ are the average 
currents corresponding to two localized states of the double-dot,
$\Delta I=I_2-I_1$, $S_I$ is the low frequency spectral density of 
the detector shot noise, and the detector nonideality is described 
by the extra dephasing due to interaction with an ``untrackable'' 
environment $\gamma =\Gamma - (\Delta I)^2/4S_I$, where 
$\Gamma$ is the dephasing rate in the conventional approach (after
ensemble averaging). In particular, the quantum point contact (QPC)  
can be an ideal detector, $\gamma =0$ (see, e.g.\ Ref.\ 
\cite{Gurvitz}), 
while the single-electron transistor (SET) in a typical operation point
is a significantly nonideal detector, $\gamma \sim \Gamma$ \cite{Shnirman}. 

        Equations (\ref{1})--(\ref{2}) allow us to calculate the evolution
of the system density matrix if the detector output $I(t)$ is known from
the experiment. They can be also used for the 
simulation, then the term $[I(t)-I_0]$ should be replaced  
with $[\Delta I (\sigma_{22}-\sigma_{11})/2 +\xi (t)]$ where the random
process $\xi (t)$ has zero average and $S_\xi = S_I$. (We 
use the Stratonovich formalism for stochastic equations.) 

        Figure 1 shows the result of such a simulation for a slightly
nonideal detector, $\gamma =0.1 \Gamma$, in the case when the evolution
starts from the maximally mixed state, $\sigma_{11}=\sigma_{22}=0.5$,
$\sigma_{12}=0$. One can see that $\sigma_{12}$ gradually appears 
during the measurement, eventually leading to well-pronounced quantum 
oscillations.
In the case $\gamma =0$ the density matrix becomes almost pure after
a sufficiently long time. This gradual purification can be interpreted 
as being due to the gradual acquiring of information about the system. 
        The detector nonideality, $\gamma \neq 0$, causes decoherence
and competes with the purification due to measurement.

        In contrast to QPC, the SET as a detector directly affects 
the two-level system asymmetry $\varepsilon$ because of the fluctuating 
potential $\phi (t)$ of SET's central island. Since there is typically 
a correlation between fluctuations of $I(t)$ and $\phi (t)$ \cite{Kor-noise},
we should add into Eq.\ (\ref{2}) the term $ i\sigma_{12} K 
[I(t)-(\sigma_{11} I_1 + \sigma_{22} I_2)]=i\sigma_{12}K\xi (t)$ 
where $K=(d\varepsilon /d\phi) S_{\phi I}/S_I\hbar$. 
This allows the partial recovery of coherence, so that 
$\gamma =\Gamma - (\Delta I)^2/4S_I - K^2 S_I/4$. The average asymmetry 
$\varepsilon$ should be also renormalized to account for the backaction 
of $\bar\phi$ shift. 

        To observe the density matrix purification experimentally,
it is necessary to record the detector output with sufficiently 
wide bandwidth, $\Delta f \gg \Gamma$ (possibly, $\Delta f \sim 10^9$ Hz),
and plug it into Eqs.\ (1)--(2). Calculations will show the development 
of quantum oscillations with precisely known phase. Stopping 
the evolution by rapidly raising the barrier ($H\rightarrow 0$) 
when $\sigma_{11}\simeq 1$ and checking that the system is really 
localized in the first state, it is possible to verify the 
presented results.

\begin{figure}
\centerline{
\epsfxsize=3.0in
\epsfbox{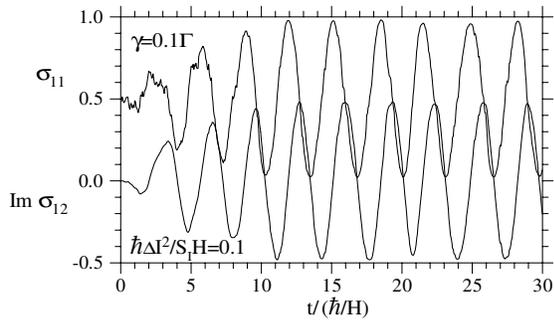}}
\caption{Gradual purification of the two-level system density matrix 
$\sigma (t)$ in a course of continuous measurement.}
\label{fig1}
\end{figure}

	The potential application in quantum computing is 
the fast initialization of the qubit state (not requiring relaxation 
to the ground state) after the intermediate measurements.

% References

\end{document}